\documentclass[10pt,twoside,twocolumn,english,aps,prl,superscriptaddress,notitlepage,showkeys]{revtex4-2}
\usepackage[utf8]{inputenc}
\usepackage{color,xcolor,soul}
\usepackage{amstext,amsfonts,amsmath,amssymb}
\usepackage{graphicx}
\usepackage[unicode=true,pdfusetitle, bookmarks=false, breaklinks=true,pdfborder={0 0 0},pdfborderstyle={},backref=false,colorlinks=true]{hyperref}
\hypersetup{colorlinks=true,citecolor=black,linkcolor=black,urlcolor=black,pagebackref=true}

\begin{document}

\newcommand{\deleted}[1]{}
\newcommand{\added}[1]{#1}
\renewcommand{\paragraph}[1]{}

\title{Coexisting Charge-Ordered States with Distinct Driving Mechanisms \\in Monolayer VSe$_2$\smallskip{}}

\author{Rebekah Chua}
\thanks{These authors contributed equally to this work}
\affiliation{Department of Physics, National University of Singapore, Singapore 117542, Singapore}
\author{Jans Henke}
\thanks{These authors contributed equally to this work}
\affiliation{Institute for Theoretical Physics Amsterdam and Delta Institute for Theoretical Physics, University of Amsterdam, Amsterdam 1098XH, The Netherlands}
\author{Surabhi Saha}
\affiliation{Department of Physics, Indian Institute of Science, Bangalore 560012, India}
\author{Yuli Huang}
\affiliation{Department of Physics, National University of Singapore, Singapore 117542, Singapore}
\affiliation{Joint School of National University of Singapore and Tianjin University, Binhai New City, Fuzhou 350207, China}
\author{Jian Gou}
\affiliation{Department of Physics, National University of Singapore, Singapore 117542, Singapore}
\author{Xiaoyue He}
\affiliation{Department of Physics, National University of Singapore, Singapore 117542, Singapore}
\affiliation{Songshan Lake Materials Laboratory, Dongguan, Guangdong 523808 China}
\author{Tanmoy Das}
\affiliation{Department of Physics, Indian Institute of Science, Bangalore 560012, India}
\author{Jasper van Wezel}
\email{j.vanwezel@uva.nl}
\affiliation{Institute for Theoretical Physics Amsterdam and Delta Institute for Theoretical Physics, University of Amsterdam, Amsterdam 1098XH, The Netherlands}
\author{Anjan Soumyanarayanan}
\email{anjan@nus.edu.sg}
\affiliation{Department of Physics, National University of Singapore, Singapore 117542, Singapore}
\affiliation{Institute of Materials Research \& Engineering (IMRE), A*STAR (Agency for Science, Technology and Research), Singapore 138634, Singapore}
\author{Andrew T.S. Wee}
\email{phyweets@nus.edu.sg}
\affiliation{Department of Physics, National University of Singapore, Singapore 117542, Singapore}

\begin{abstract}
Thinning crystalline materials to two dimensions (2D) creates a rich playground for electronic phases, including charge, spin, superconducting, and topological order. Bulk materials hosting charge density waves (CDWs), when reduced to ultrathin films, have shown CDW enhancement and tunability. However, charge order confined to only 2D remains elusive. Here we report a distinct charge ordered state emerging in the monolayer limit of $1T$-VSe$_2$. Systematic scanning tunneling microscopy experiments reveal that bilayer VSe$_2$ largely retains the bulk electronic structure, hosting a tri-directional CDW. However, monolayer VSe$_2$ -- consistently across distinct substrates -- exhibits a dimensional crossover, hosting two CDWs with distinct wavelengths and transition temperatures. Electronic structure calculations reveal that while one CDW is bulk-like and arises from the well-known Peierls mechanism, the other is decidedly unconventional. The observed CDW-lattice decoupling and the emergence of a flat band suggest that the new CDW \added{could} arise from enhanced electron-electron interactions in the 2D limit. These findings establish monolayer-VSe$_2$ as \deleted{the first}\added{a} host of coexisting charge orders with distinct origins, \deleted{opening the door to}\added{and enable the} tailoring \added{of} electronic phenomena \emph{via} emergent interactions in 2D materials.
\end{abstract}

\added{}

\maketitle

\paragraph{Motivation}
Charge order in crystalline materials typically manifests as a static modulation of electron density, known as a charge-density wave (CDW), accompanied by periodic modulations of the atomic lattice~\citep{Gruner1988}. The prototypical CDW arises in (quasi-)one-dimensional (1D) systems from the ``nesting'' of parallel Fermi surface (FS) regions connected by the CDW propagation vector $Q_{{\rm CDW}}$. As real materials do not exhibit perfect nesting, CDW formation is supported by either electron-phonon coupling (EPC), other collective excitations, or electron-electron interactions~\citep{Johannes2008,Zhu2015,Feng2015,Flicker2015,Kogar2017,Henke2020}.
In layered materials, CDWs often exist in proximity to other ordered phases, \emph{e.g.} superconductivity and magnetism~\citep{Chen2016}, due to a precarious balance between competing interactions. Approaching the two-dimensional (2D) limit enhances the potential for such interplay~\citep{Steinke2020}, while providing new knobs to tune electronic phases, such as electric fields and strain~\citep{Novoselov2005b,Li2015a,Tsen2015,Gao2018}. Notably, electron-electron interactions in the 2D limit are expected to induce competition among different CDW driving mechanisms as well as other ordered states ~\citep{Abram2017, Butler2021, Lopes2021}. In practice, however, a crossover towards electronic charge order driven by dimensional reduction remains to be discovered.
\noindent %

\begin{figure*}
\begin{centering}
\includegraphics[width=5.2in]{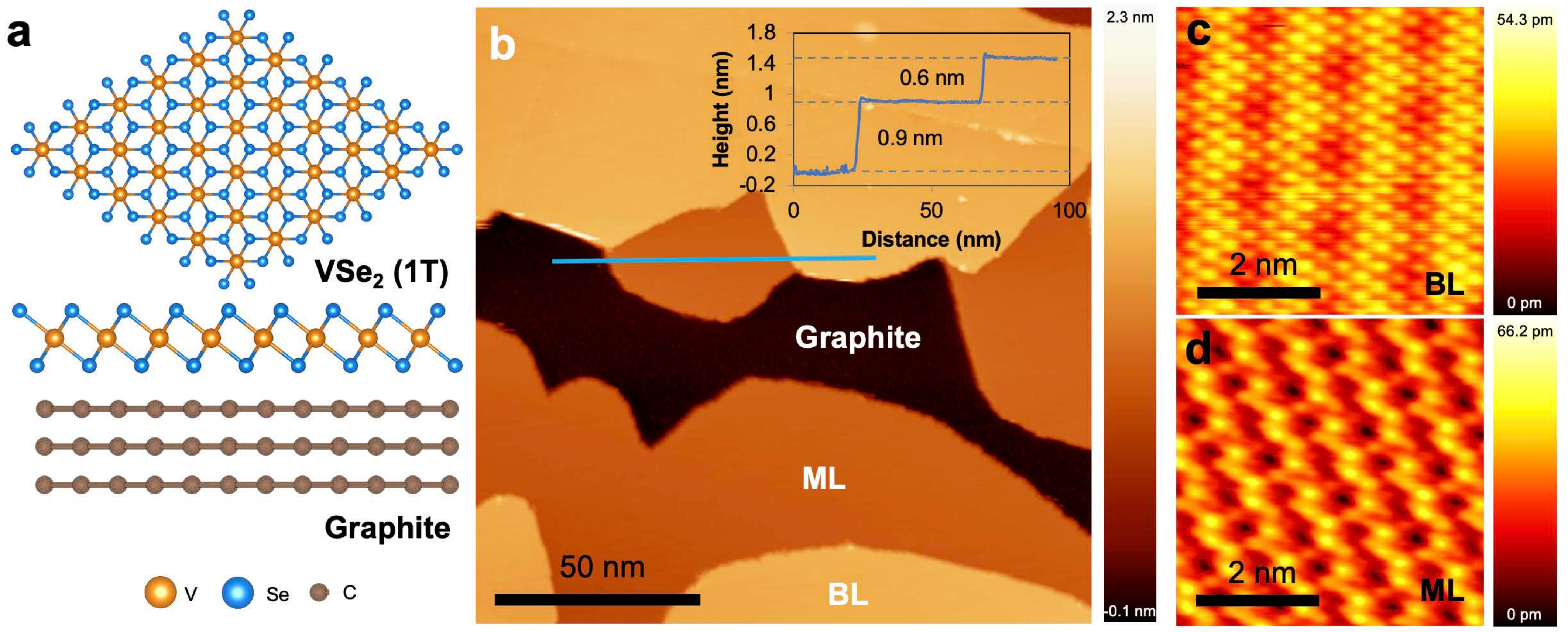}
\end{centering}
\caption[Materials \& Methods]{\textbf{Crystal structure and STM Imaging of Ultrathin VSe$_2$.} 
\textbf{(a)} Top view (top) and side view (bottom) of the atomic structure of $1T$-VSe$_2$ on graphite substrate.
\textbf{(b)} Large field-of-view STM topograph ($150 \times 150$\,nm$^2$, bias voltage, $V_{\rm tip} = 0.7$~V, current setpoint, $I_{\rm set} = 100$~pA) of MBE-grown epitaxial VSe$_2$ on HOPG substrate (see Methods for details). The average thickness of 1.5 VSe$_2$ layers resulted in monolayer (ML) and bilayer (BL) VSe$_2$ regions identified within the image. Inset shows the height profile across layers, with the respective step heights indicated.
\textbf{(c-d)} Atomically resolved STM topographs ($5\times 5$\,nm$^2$, $V_{\rm tip} = -0.2$~V, $I_{\rm set} = 200$~pA) of BL (c) and ML (d) VSe$_2$ at 78~K. Both topographs show a hexagonal lattice with visibly distinct superstructures. 
\label{fig:stm1}}
\end{figure*}

\paragraph{TMDC Lit Review}
Transition metal dichalcogenides (TMDCs) are well-studied hosts of conventional and unconventional CDWs~\citep{Rossnagel2011,Chen2016,Feng2015,Flicker2015,Kogar2017}. The tunability of CDWs in the ultrathin limit of several TMDCs is particularly relevant to practical electronic applications~\citep{Tsen2015,Xi2015,Li2015a,Barja2016,Gao2018}. $1T$-VSe${2}$ is a prototypical metallic TMDC with layered hexagonal crystal structure (\ref{fig:stm1}a). Bulk $1T$-VSe$_{2}$ is paramagnetic, with a three-dimensional (3D) FS. Below temperature $T_{{\rm CDW}}^{{\rm bulk}}\sim 110$~K, it hosts a triple-$Q$ (triangular) CDW with 3D character. The CDW periodicity $\lambda_{{\rm CDW}}\simeq4a\times4a\times3c+\delta$ is commensurate with the in-plane lattice constant $a$, but incommensurate with the inter-layer distance $c$, and corresponds to a weakly nested FS region, supported by a structured EPC~\citep{Strocov2012,Henke2020}. For thicknesses below 20~nm, the FS of $1T$-VSe$_{2}$ transitions to 2D character, while maintaining triple-$Q$, $4a$ CDW order~\citep{Pasztor2017}.

\paragraph{ML-VSe2 Lit Review}
Meanwhile, monolayer (ML)-VSe$_{2}$, grown epitaxially in several recent works, purportedly hosts a ground state with concomitant charge and spin orders, the nature of which is controversial~\citep{Bonilla2018,Duvjir2018,Chen2018}. First, while some claim $4a$ CDW to be absent even at low temperatures~\citep{Bonilla2018,Chen2018}, others indicate its persistence to well above room temperature~\citep{Duvjir2018}. Second, several works report incommensurate superstructures with varying periodicities, viz. $\sqrt{3}a \times 2a$, $\sqrt{3}a \times \sqrt{7}a$, and $\sim2a \times 3a $~\citep{Duvjir2018,Bonilla2018,Chen2018, Ly2020}, whose purported origins vary from structural distortions to nested CDWs. The relation of all these superstructures -- identified \emph{via} electronic density distributions over small real space regions -- to any long-ranged charge order remains unclear. Finally, magnetism is suggested to emerge in ML-VSe$_{2}$ despite its absence in the bulk~\citep{Bonilla2018,Wong2019}, but both its existence and interplay with charge order are actively debated~\citep{Coelho2019,Chua2020}. Disentangling these apparently conflicting observations is paramount to revealing the true nature of charge order in ML-VSe$_{2}$, its driving mechanism, and its ramifications on other phases. This requires a controlled and systematic study of the CDW under varying thermodynamic conditions.

\paragraph{Results Summary}
Here we report a comprehensive experimental and theoretical investigation of charge order in ultrathin epitaxial $1T$-VSe$_{2}$. Scanning tunnelling microscopy (STM) experiments show that while the CDW in BL-VSe$_{2}$ is closely related to that in bulk, charge order in ML-VSe$_{2}$ is qualitatively different. By systematically varying substrates, film thickness, and temperature, we find that ML-VSe$_{2}$ consistently hosts two unidirectional (single-$Q$) CDWs with periods $4a$ and $2.8a$, with strikingly distinct phenomenologies. Band structure calculations elucidate that while the $4a$ CDW is stabilized by conventional FS nesting and EPC, the $2.8a$ CDW cannot be explained by such mechanisms. Instead, we find the $2.8 a$ instability to originate from a flat band region, wherein electron-electron interactions are \added{expected to be} strongly enhanced. Our results establish ML-VSe$_{2}$ as \deleted{the first material hosting}\added{a host of} coexisting CDWs with distinct driving mechanisms, demonstrating the potential of correlations for tuning electronic phases in the 2D limit.

\section{Results and discussion}
\subsection*{STM Imaging Experiments\label{sec:Expt-Results}}
\paragraph{Materials \& Expt Setup}
Thin films of VSe$_{2}$ were grown using molecular beam epitaxy (MBE) on highly oriented pyrolytic graphite (HOPG) and MoS$_{2}$ substrates under ultrahigh vacuum conditions (see Methods). Both substrates are known to stabilize the $1T$ polymorph of VSe$_{2}$~\citep{Bonilla2018} whose crystal structure is shown in \ref{fig:stm1}a. The films were characterized \emph{in-situ} using STM over temperatures of 77-200~K (see Methods). As shown in \ref{fig:stm1}b, controlled growth of an average thickness of 1.5 layers resulted in the formation of both ML- and BL-VSe$_{2}$ regions (on HOPG) within fields-of-view accessible to STM imaging. Topographic characterization of a terraced region at 78~K (\ref{fig:stm1}b: inset) reveals step heights of 0.9\,nm and 0.6\,nm for the first and second VSe$_{2}$ layers respectively, in line with values reported previously~\citep{Duvjir2018}. 

\paragraph{CDWs Overview}
\ref{fig:stm1}c-d display atomic resolution topographs obtained in the BL and ML regions, respectively. As expected, both cases show the expected hexagonal arrangement of atoms with lattice constant, $a\simeq 0.34$\,nm~\citep{Bonilla2018, Duvjir2018}. Meanwhile, the atomic-scale superstructures seen on ML- and BL-VSe$_{2}$ appear starkly different. For BL-VSe$_{2}$ (\ref{fig:stm1}c), the superstructure is tri-directional, \emph{i.e.} it manifests along all three lattice directions with a single lengthscale. The overall phenomenology is remarkably similar to that of the triple-$Q$ CDW reported in bulk and thinned 1T-VSe$_{2}$ crystals~\citep{Pasztor2017}. 
In contrast, for ML-VSe$_{2}$ (\ref{fig:stm1}d), the superstructure appears unidirectional, and has multiple lengthscales, consistent with recent results reported by other groups~\citep{Bonilla2018, Duvjir2018}. Crucially, complementary imaging of the ML using non-contact atomic force microscopy under similar conditions shows no corrugations beyond those of the atomic lattice (see Supporting Information (SI) \S S2), which rules out structural distortions. Therefore, we conclude that the superstructures observed in STM imaging of ML-VSe$_{2}$ must be of electronic origin, and putatively regard them as CDWs.    

\begin{figure*}
    \begin{centering}
    \includegraphics[width=5.4in]{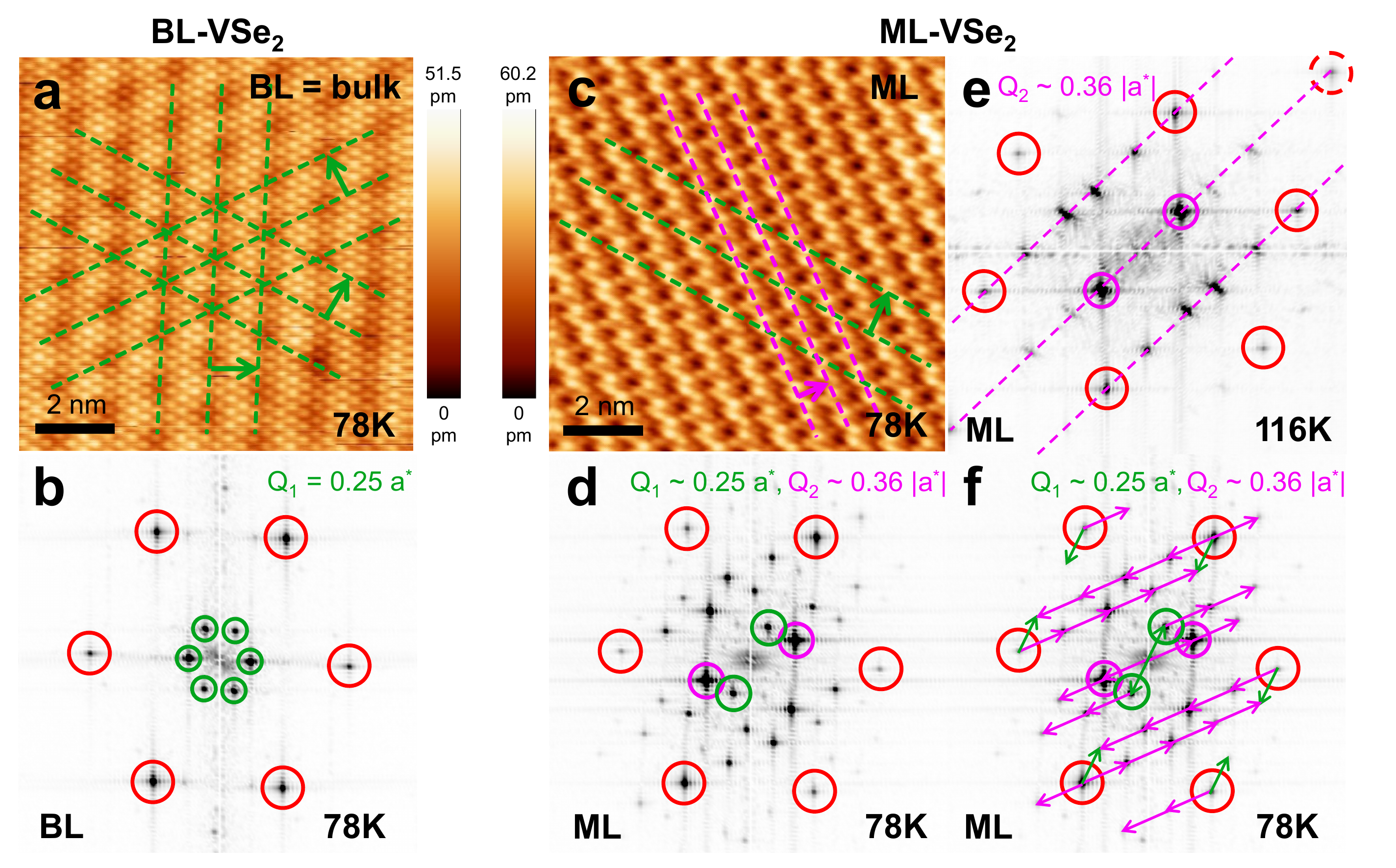}
    \end{centering}
    \caption[ML vs. BL Comparison]{\textbf{Comparison of CDWs in BL and ML-VSe$_2$.} 
    \textbf{(a-d)} STM topographs (a, c: $10 \times 10$\,nm$^2$, $V_{\rm tip} = -0.2$~V, $I_{\rm tip} = 200$~pA ) and their respective Fourier Transforms (FTs: b, d) acquired \added{at 78~K} on BL- (a-b) and ML- (c-d) VSe$_2$ \deleted{at 78~K}\added{from adjacent terraces with no observable grain boundary}. Dashed colour-coded lines in (a, c) represent the real space CDW wavefronts, and corresponding circles in (b, d) denote the respective CDW wavevectors $Q_1$ (b, d: green) and $Q_2$ (d: magenta), whose magnitudes are indicated in reciprocal lattice units (rlu). Red circles denote atomic Bragg peaks in all FT images. 
    \textbf{(e)} FT of STM topograph  acquired on ML-VSe$_2$ at 116~K. Magenta circle denotes the $Q_2$ peak, while the $Q_1$ peak is absent. Dashed lines indicate the orientation of $Q_2$ and its harmonics with respect to the Bragg peak (red circle). 
    \textbf{(f)} Annotated FT of ML-VSe$_2$ at 78 K (c.f. data in d). Green ($Q_1$) and magenta ($Q_2$) circles identify the primary CDW peaks. Color-coded arrows indicate the positions of harmonics with respect to primary and Bragg peaks. All peaks can be accounted for this way.
    \label{fig:stm2}}
\end{figure*}

\paragraph{Disentangling ML CDWs}
In light of conflicting reports on the CDW phenomenology in ultrathin VSe$_{2}$, we systematically examine in \ref{fig:stm2} the Fourier space modulations from larger STM topographs obtained for both BL and ML cases. For BL-VSe$_{2}$, \ref{fig:stm2}b shows the Fourier transform (FT) of a typical STM topograph. Here we find prominent peaks at $Q_1 \simeq 0.25 a^{*}$ (green circles), where $a^{*}$ is the magnitude of the reciprocal lattice vector, with $\mathcal{C}_6$ symmetry, \emph{i.e.} along all three Bragg directions. \added{Meanwhile, the anisotropy of Bragg peak intensities may indicate either local uniaxial strain within the sample, or asymmetry in the tip shape. Regardless, } these observations are \deleted{fully} consistent with the triple-$Q$, $4 a$ CDW reported in bulk and thinned $1T-$VSe$_{2}$ crystals~\citep{Pasztor2017}. In contrast, the FT for ML-VSe$_{2}$ shown in \ref{fig:stm2}d appears more complex, with only $\mathcal{C}_2$ symmetry present.  Firstly, c.f. the BL, the ML shows the persistence of the $Q_1 \simeq 0.25 a^{*}$ peak (green circle) along a single Bragg direction, corresponding to a single-$Q$, $4 a$ CDW. Secondly, the most prominent Fourier peak for the ML is seen at $Q_2 \simeq 0.36 a^{*}$ at an angle $\theta_{12} \sim 30^{\circ}$ relative to the Bragg direction (magenta circle). As shown in \ref{fig:stm2}f, a careful inspection of the FT for the ML suggests that all remaining Fourier peaks can be assigned to higher harmonics or Bragg reflections of $Q_1$ and $Q_2$, including previously reported multiplet superstructures~\citep{Duvjir2018,Chen2018,Ly2020}. While such superstructures may, in principle, be identified with several distinct wavelengths over small topographic regions, such identifications are not consistent over length scales above 5~nm in any of the reported data~\citep{Duvjir2018,Chen2018,Ly2020}. Instead, we propose that these apparent supercells are merely the result of superposing two single-$Q$ CDWs, one of which is aligned away from a high-symmetry direction and also incommensurate with the atomic lattice.

\subsection*{Temperature Dependence\label{sec:TDepSTM}}

\paragraph{Two distinct CDWs}
To further establish the character of CDW(s), we studied the evolution of CDW peaks in BL- and ML-VSe$_2$ with temperature, across both substrates. Notably, the FT of ML-VSe$_{2}$ recorded at higher temperatures (\ref{fig:stm2}e) reveal only a single Fourier modulation with magnitude $Q_2$, as well as its harmonics and reflections. This further evidences the presence of only two principal CDWs -- $Q_1$ and $Q_2$ -- and suggests that they may have independent origins. At the same time, the slight thermal variation in the direction of $Q_2$ with respect to the lattice shows that the $Q_2$ CDW is not strongly coupled to the lattice. It also suggests a potential interplay between the two CDWs, which may lower the energetic cost of the charge ordered state when harmonics and reflections of $Q_2$ are connected by $Q_1$ (\ref{fig:stm2}f). Meanwhile, both BL- and ML-VSe$_2$ grown on MoS$_2$ substrate (see SI \S S1) exhibit identical CDW phenomenology to their counterparts grown on HOPG (\ref{fig:stm2}), \deleted{suggesting limited influence}\added{limiting the potential role} of substrate-induced strain effects \added{in driving CDW formation}. 

\begin{figure*}
    \begin{centering}
    \includegraphics[width=5in]{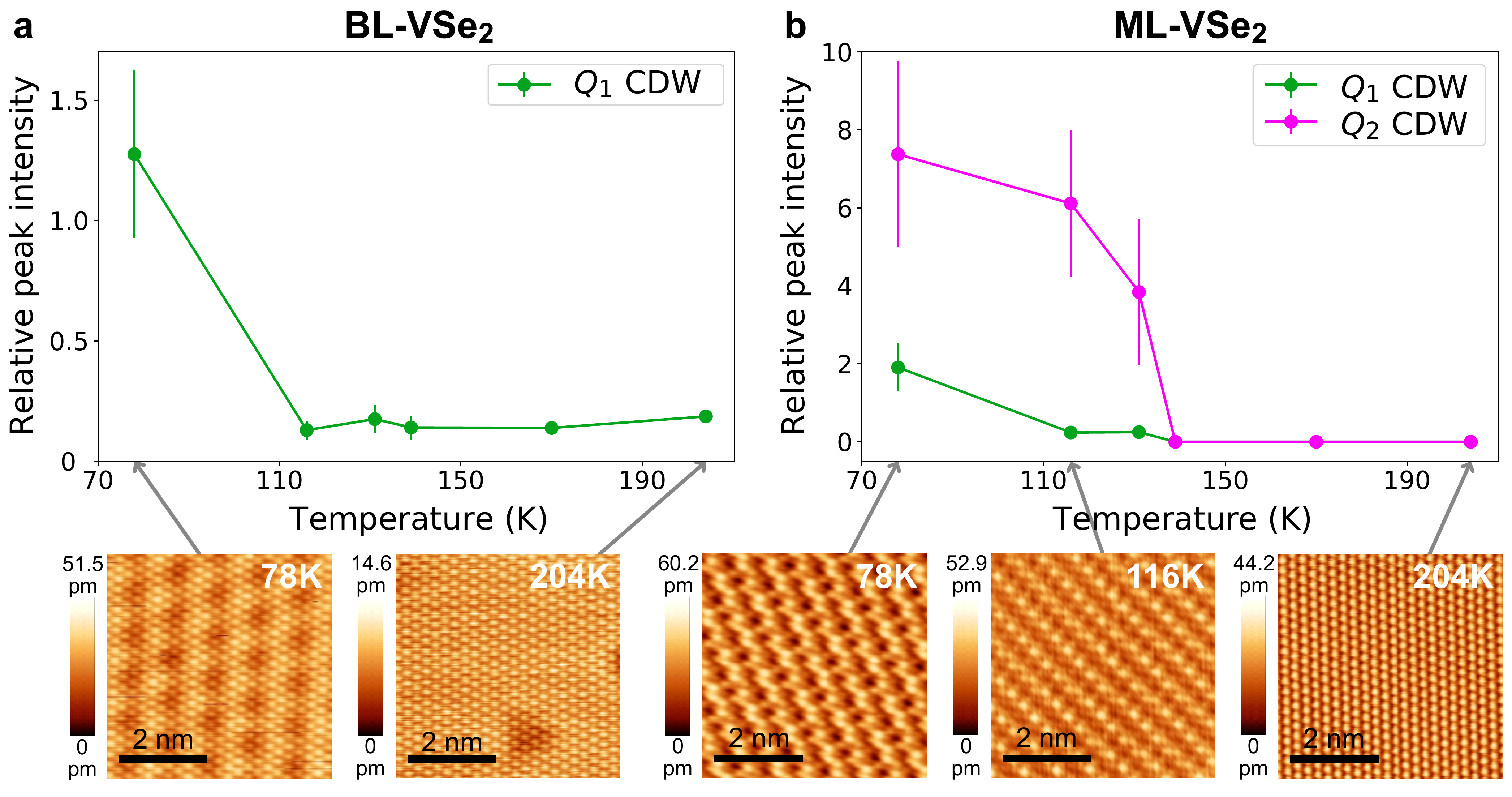}
    \end{centering}
    \caption[$T$-Evolution of CDW Intensities]{\textbf{Temperature Dependence of CDW Intensities in ultrathin VSe$_2$.} 
    Thermal evolution of intensities of the CDW peaks, normalized to the averaged intensities of the six primary Bragg peaks at that temperature, as measured from FTs of $10\times 10$\,nm$^2$ STM topographs acquired on (a) BL- and (b) ML-VSe$_2$ respectively (dataset in SI \S S3, Fig. S3). Error bars show the standard deviation, incorporating the variance in Bragg and CDW peak intensities at each temperature. 
    Insets show STM topographs at selected temperatures for BL (left) and ML (right) respectively. \label{fig:stm3}}
\end{figure*}

\paragraph{CDW Intensity Variation}
The thermal evolution of the CDW intensity in STM topographs is an established thermodynamic marker of the CDW transition~\citep{Arguello2014, Chatterjee2015}. In \ref{fig:stm3}, we show representative STM topographs for ML-VSe2 on HOPG for different temperatures (extended dataset in SI \S S3). While the data were recorded over varying fields-of-view, we emphasize that, within our experiments, none of the CDWs exhibit any macroscopic spatial variation across atomically smooth regions. For ease of comparison, the CDW peak intensities plotted in \ref{fig:stm3} are normalized to the corresponding Bragg peak intensities for each STM topograph. Consistently across BL- and ML-VSe$_{2}$, we find that the intensity of $Q_1$ ($4a$ CDW) drops sharply at $\sim 110$~K to a negligible magnitude, consistent with the thermal evolution of its bulk counterpart~\citep{Pasztor2017}. 
The small, finite magnitude of $Q_1$ in BL-VSe$_{2}$ at higher temperatures likely arises from small CDW pockets near defects, similar to defect-pinned CDWs at $T\gg T_{\rm CDW}$ reported in other TMDCs~\citep{Arguello2014, Chatterjee2015}. Meanwhile, for ML-VSe$_2$, the intensity of $Q_2$ -- in sharp contrast to $Q_1$ -- remains sizable well above $\sim 110$~K, and drops to nearly zero at $\sim 140$~K. Finally, no CDW signatures are observed in the 204~K topographs (\ref{fig:stm3}c,g), precluding the persistence of either CDW to room temperature~\citep{Duvjir2018}.

\paragraph{CDWs Expt Summary}
Overall, our systematic analysis sheds much-needed light on the presence, character, and robustness of putative charge order in ML-VSe$_2$ in view of conflicting reports in literature \citep{Zhang2017,Duvjir2018,Chen2018,Chen2020,Duvjir2021}. First, our AFM-STM comparison confirms the purely electronic (CDW) origin of all observed superstructures on ML- and BL-VSe$_2$ (c.f. \citep{Duvjir2018}). Second, $T$-dependent experiments conclusively establish the presence of two, and only two, independent single-$Q$ CDWs in ML-VSe$_{2}$ -- $Q_1 \simeq 0.25 a^{*}$ (\emph{i.e.} $\lambda_1 \simeq 4 a$) and $Q_2 \simeq 0.36 a^{*}$ (\emph{i.e.} $\lambda_2 \simeq 2.8 a$), respectively. The $Q_1$ CDW is identical in magnitude, orientation, and transition temperature to the triple-$Q$ CDW observed in BL-VSe$_{2}$, and to (the in-plane projection of) the CDW reported in bulk crystals. Meanwhile, the  $Q_2$ CDW persists at temperatures well beyond $Q_1$ and exhibits thermal variations in its orientation with respect to the atomic lattice. Finally, the observed consistency of $Q_1$ and $Q_2$ across distinct substrates (c.f.\citep{Chen2020,Duvjir2021}), and of BL-VSe$_2$ with bulk (c.f.\citep{Zhang2017}), strongly constrain the potential influence of substrate-induced strain effects on the CDW characteristics reported here. To understand the origin of this observed dichotomy in CDW characteristics within the same material, we conduct a detailed examination of the electronic structure of ultrathin VSe$_{2}$.

\subsection*{Band Structure Calculations\label{sec:BandStructure}}

\begin{figure*}
    \begin{centering}
    \includegraphics[width=0.8\linewidth]{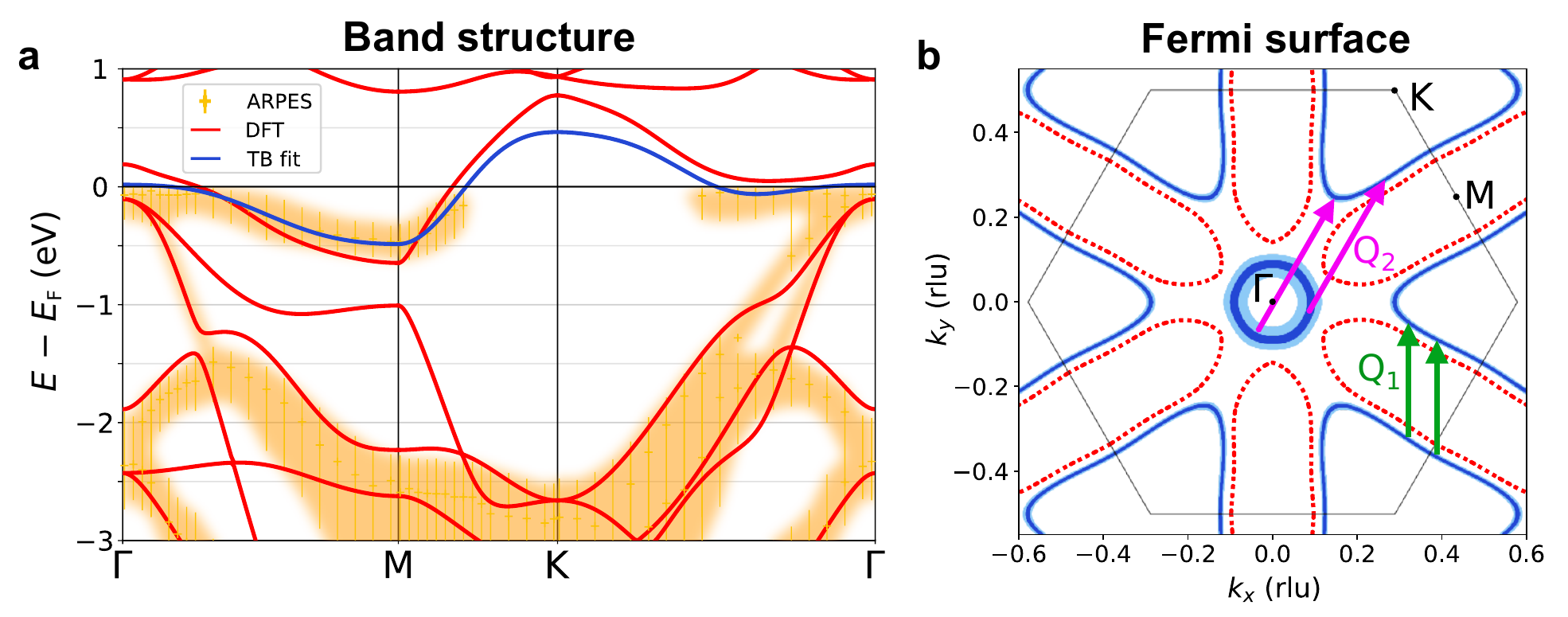}
    \end{centering}
    \caption[ML-VSe$_2$ Band Structure]{\textbf{Electronic Structure of ML-VSe$_2$.}
    \textbf{(a)} Electronic band structure of ML-VSe$_2$ obtained from DFT calculations (red line), compared to published ARPES measurements of the spectral function for epitaxially grown ML-VSe$_2$ at $ T = 170$~K (shaded yellow, crosses: peak positions, lines: full width at half maximum). 
    The linewidth of the experimental data greatly exceeds the experimental resolution~\citep{Feng2018}.
    Blue line is a tight-binding (TB) fit to the ARPES-measured, near-$E_{\rm F}$ band structure, where  $E_{\rm F}$ is the Fermi energy. 
    \textbf{(b)} Fermi surface (FS) of ML-VSe$_2$, obtained from the TB fit in (a) by plotting states within $\pm 1$\,meV (dark blue) and $\pm 10$\,meV ($\approx k_{b}T$ for $T=100$\,K, light blue) of $E_{\rm F}$. Dotted red line shows the DFT FS, which qualitatively deviates from the TB fit. Hexagon shows the Brillouin zone (BZ), and the arrows indicate FS regions visually appearing to be nested by the experimentally determined CDW wavevectors. \label{fig:BandStructure}}
\end{figure*}

\paragraph{DFT \& Atomic structure}
Density functional theory (DFT) calculations were performed to investigate the atomic and electronic structure of ultrathin $1T$-VSe$_{2}$ using the Vienna Ab-initio Simulation Package (VASP, see Methods)~\citep{Kresse1996b}. ML-VSe$_{2}$ was simulated by requiring the interlayer distance to be 25~{\AA}, and relaxing a $4\times 4$ atomic supercell structure, both with and without the symmetry constraints of the underlying $P\bar{3}m1$ space group~\citep{Li2014}. In both cases, the resulting lattice is purely hexagonal, and free of any structural distortions (c.f.~\citep{Duvjir2018}). This further points to the electronic origin of superstructures observed in ML-VSe$_{2}$, in line with our experimental findings. Subsequently, the electronic structure was computed, both with and without including spin polarization. The resulting energies are nearly equal for both cases. This suggests, in conjunction with the absence of spin splitting in angle-resolved photoemission spectroscopy (ARPES) results~\citep{Duvjir2018,Feng2018,Chen2018,Wong2019,Umemoto2019,Coelho2019,Biswas2020}, that magnetic order, even if present in ML-VSe$_{2}$, is unlikely to play a significant role in the energetics of charge ordered states.

\paragraph{Band Structure Comparison}
The DFT band structure (\ref{fig:BandStructure}a) is broadly in agreement with the ARPES spectral function measured for ML-VSe$_{2}$~\citep{Feng2018}. The data in ref.~\citep{Feng2018} provides a valuable benchmark given its high quality, large momentum range, and qualitative agreement with other ARPES reports, including data acquired on our samples (SI \S S4, Fig. S4)~\citep{Wong2019}. Both techniques find a single band of predominantly $d$-orbital character crossing the Fermi energy $E_{\mathrm{F}}$. Previous works have emphasised the importance of the nesting of the sides of the FS lobes at the BZ edge~\citep{Duvjir2018, Chen2018, Trott2020}. The DFT electronic structure, however, underestimates $\mathbf{k}_F$ along $\mathrm{M - K}$ and suggests a ``nesting vector'' along $\mathbf{a}^*$ of length 0.21\,rlu. This falls short of the vector extracted from ARPES data ($0.54\pm0.04$\,\AA$^{-1}$), which corresponds to $0.25\pm0.02$\,rlu~\citep{Feng2018, Coelho2019}. The DFT band along $\Gamma-\mathrm{M}$ also appears more dispersive than that in ARPES, while along $\Gamma-\mathrm{K}$ the DFT band is higher (50-200\,meV) than the magnitude expected from the high photoelectron count around $\Gamma$~\citep{Duvjir2018,Feng2018,Chen2018,Wong2019,Umemoto2019,Coelho2019}. These discrepancies are likely due to the inability to duly account for electronic correlations~\citep{Biswas2020}. As a result, our \emph{ab initio} calculations may not capture the electronic structure near $E_F$ with sufficient quantitative accuracy to describe CDW energetics. 

\begin{figure*} 
    \begin{centering}
    \includegraphics[width=0.92\linewidth]{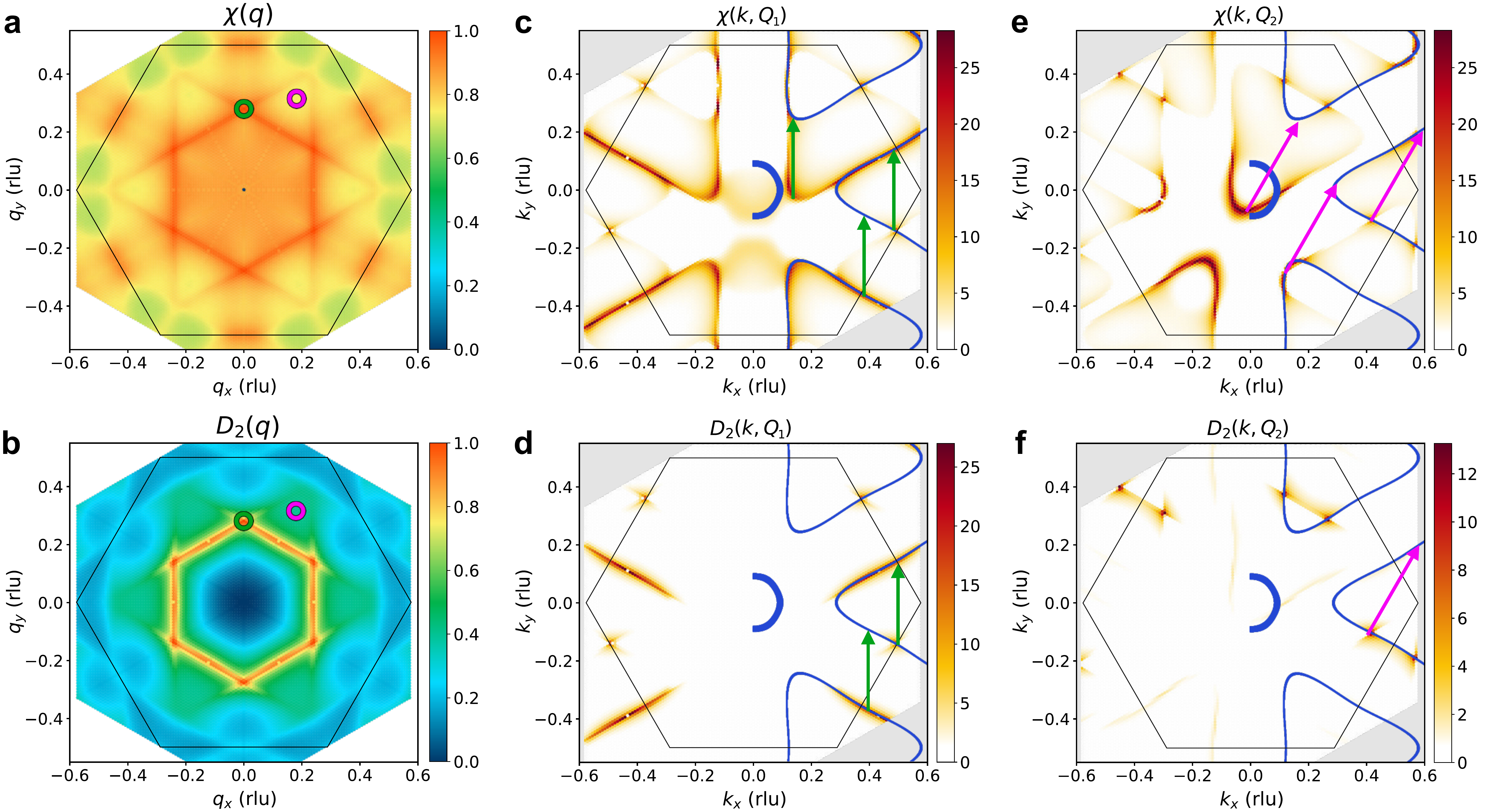}
    \end{centering}
    \caption[Susceptibility Calculations]{\textbf{Momentum Space Diagnostics of ML-VSe$_2$ CDWs.} 
    \textbf{(a)} The normalised Lindhard susceptibility $\chi$ and \textbf{(b)} structured electronic susceptibility $D_2$ derived from the TB band structure. Circles highlight the positions of the susceptibility maximum $(0, 0.28)\approx \mathbf{Q}_1$ (green) and experimentally determined $\mathbf{Q}_2=(0.182,0.315)$ (magenta), with the latter being located on an intensity plateau. \textbf{(c-f)} Diagnostics indicating the BZ regions contributing to the two susceptibilities $\chi$ and $D_2$, given a chosen wavevector $\mathbf{Q}_1$ (c,d) or $\mathbf{Q}_2$ (e,f). Blue lines denote FS contours in the right half of the images, while black hexagons indicate the BZ. $\mathbf{Q}_1$ (green) and $\mathbf{Q}_2$ (magenta) arrows indicate the regions with prominent contributions to the susceptibilities as deduced from the diagnostics. \label{fig:chiD2}}
\end{figure*}

\paragraph{TB Band Structure}
We therefore complement the DFT calculation with a tight-binding (TB) fit to the ARPES data in ref.~\citep{Feng2018} (see Methods), the results of which are compared to the DFT in \ref{fig:BandStructure}. In agreement with reported ARPES spectra, the TB fit shows a flat band region around the $\Gamma$-point, an indicator of strong correlations. The difference in topology between the DFT and TB FS (Fig. \ref{fig:BandStructure}b) is due to the proximity of a van Hove singularity to $E_F$~\citep{Feng2018}. Overlaying the CDW vectors extracted from our STM data onto the FS visually suggests that $Q_1$ corresponds to nesting between the sides of neighbouring triangular FS pockets at the BZ edge, while $Q_2$ connects the flat-band region around $\Gamma$ to the pocket corners around K.

\subsection*{Nesting and Correlated Instabilities\label{sec:Susc}}

\paragraph{Susceptibility Calc}
A conventional CDW instability at wavevector $\mathbf{Q}_{\rm CDW}$ results from a maximum in its electronic susceptibility $D_{2}(\mathbf{q})$ for $\mathbf{q}=\mathbf{Q}_{\rm CDW}$~\citep{Doran1978a,Johannes2008}. In the weak electron-phonon coupling (EPC) limit (see Methods), $D_{2}(\mathbf{q})$ can be expressed as~\citep{Doran1978a}: 
\begin{align}
D_{2}(\mathbf{q})=-\sum_{\mathbf{k}\in\text{BZ}}\left|g_{\mathbf{k},\mathbf{k+q}}\right|^{2}\frac{f(E_{\mathbf{k}})-f(E_{\mathbf{k+q}})}{E_{\mathbf{k}}-E_{\mathbf{k+q}}+i\delta}.
\end{align}
Here, $f(E)$ is the Fermi-Dirac function, $E_{\mathbf{k}}$ is the bare (non-renormalized) electronic dispersion, and $\delta$ is a small regulator (0.1\,meV in this work). The EPC matrix elements, $\left|g_{\mathbf{k},\mathbf{k+q}}\right|$, are often approximated to unity, resulting in the Lindhard, or bare susceptibility, $\chi(\mathbf{q})$. However, for TMDCs whose near-$E_{\rm F}$ behaviour is governed by $d$-band(s), several works have established a more realistic approximation to $\left|g_{\mathbf{k},\mathbf{k+q}}\right|$ \emph{via} the electronic band structure~\citep{Varma1979,Flicker2016,Henke2020} (see Methods). Here, we use the TB fit to calculate the bare ($\chi(\mathbf{q})$) and structured ($D_{2}(\mathbf{q})$) electronic susceptibilities, which are shown in \ref{fig:chiD2}a and b, respectively.

\paragraph{Susceptibility: $\mathbf{Q}_1$ CDW}
The green circle in \ref{fig:chiD2}a(b) indicates the maximum of the bare (structured) susceptibility, which lies at $\mathbf{Q}=(0,0.28)\approx \mathbf{Q}_1$. Its proximity to a commensurate value suggests that the corresponding CDW will lock to 0.25\,rlu ($\lambda = 4a$) due to CDW-lattice interactions~\citep{Feng2015}. Although its periodicity is the same as that of the CDW observed in bulk and BL-VSe$_2$, the FS for the ML is strictly 2D, and the parts of the FS involved in CDW formation may be different.
To elucidate the role of the FS in the observed CDWs, we plot in \ref{fig:chiD2}c, d the $\mathbf{k}$-resolved contributions to $\chi(\mathbf{q})$ and $D_{2}(\mathbf{q})$ for  $\mathbf{q} = \mathbf{Q}_1$. As anticipated in \ref{fig:BandStructure}b, the dominant contributions to $\chi(\mathbf{Q}_1)$ arise from the parallel edges of the ${\rm K}$-centred pockets, while the $\Gamma$-centred FS region plays a negligible role. The well-nested ${\rm K}$-pocket edges with opposite group velocities are therefore inherently unstable to a Peierls-like CDW. The EPC matrix elements further enhance the contribution of these $\mathbf{Q}_1$-connected regions to $D_{2}(\mathbf{Q}_1)$, thereby confirming the conventional origin of the $\mathbf{Q}_1$ CDW in ML-VSe$_2$.

\paragraph{Susceptibility: $\mathbf{Q}_2$ CDW}
In contrast, the phenomenology for $\mathbf{q} = \mathbf{Q}_2$ does not fit the conventional CDW framework. As highlighted by the magenta circles in \ref{fig:chiD2}a, b, this wavevector lies in the middle of a susceptibility plateau, and lacks a well-defined maximum. The dominant contribution to the bare susceptibility at $\mathbf{q} = \mathbf{Q}_2$ comes from the $\Gamma$-centered flat band region, with smaller contributions from the ${\rm K}$-centred pockets (see \ref{fig:chiD2}e). However, the corresponding $D_{2}(\mathbf{k},\mathbf{Q}_2)$ in \ref{fig:chiD2}f shows that the EPC matrix elements strongly suppress the intensity in these regions, and the remaining contributions are insufficient to drive the $\mathbf{Q}_2$ CDW according to an EPC-assisted Peierls scenario. While the perturbative expansion used for the structured susceptibility calculations~\citep{Doran1978a,Varma1979,Flicker2016,Henke2020} may not fully capture EPC in flat bands, that the origin of the $\mathbf{Q}_2$ CDW lies beyond the Peierls description of CDWs is consistent with its empirical characteristics, viz. varying orientation with respect to the lattice, absence in BLs (and beyond), and the lack of a discernible peak in $\chi(\mathbf{q})$.

\paragraph{StrEl Origin of $\mathbf{Q}_2$ CDW}
In the 2D limit of layered TMDCs like $1T$-VSe$_{2}$, the screening of Coulomb interactions between electrons is much reduced~\citep{Nozieres1964}. The relative importance of the unscreened interactions is further enhanced within flat bands associated with a van Hove singularity, such as at the near-$\Gamma$ region in ML-VSe$_{2}$ (\ref{fig:BandStructure})~\citep{Duvjir2018,Feng2018,Umemoto2019,Biswas2020}. Indeed, the measured linewidth, or self-energy, of the band near $E_F$ is much larger than the experimental resolution~\citep{Feng2018}, \deleted{confirming}\added{supporting} the presence of strong electronic correlations~\citep{Damascelli2003}. Such interactions can considerably renormalise electron and phonon properties, and enable CDW order at momenta that do not correspond to peaks in the conventional susceptibility 
($\chi(\mathbf{q})$ or $D_{2}(\mathbf{q})$). Indeed, such correlation-driven CDWs have been predicted to exist in TMDCs~\citep{Chen2016}, including in ML-VSe$_2$~\citep{Trott2020}, and are consistent with the unusual characteristics of the $\mathbf{Q}_2$ CDW. Crucially, a correlation-driven mechanism for the $\mathbf{Q}_2$ CDW offers the only viable explanation of its prevalence over a well-nested counterpart ($\mathbf{Q}_1$), and the complete gapping of the FS~\citep{Duvjir2018,Feng2018,Umemoto2019,Biswas2020}, despite the absence of any associated feature in susceptibility calculations based on models of non-interacting electrons. \added{Further, we conjecture that the single-$q$ character of the $\mathbf{Q}_2$ CDW, which breaks the three-fold rotational symmetry of the lattice, makes it energetically favourable for the $\mathbf{Q}_1$ CDW (nominally triple-$q$) to also order in a single-$q$ configuration. The interplay of these CDWs could be examined in future theoretical works by iteratively incorporating the resulting lattice distortions.}

\section{Conclusions\label{sec:Outlook}}
\paragraph{Results Summary}
In summary, our systematic experimental and theoretical efforts elucidate that $1T$-VSe$_{2}$ undergoes a dimensional crossover as its thickness is reduced to a single layer. While BL-VSe$_{2}$, akin to bulk, hosts a conventional triple-$Q$ CDW, ML-VSe$_{2}$ hosts two distinct single-$Q$ CDWs with contrasting characteristics. One, with $\lambda_1\,\simeq\,4 a$, behaves similarly to its BL/bulk counterpart, and arises from a weak-coupling Peierls mechanism utilizing nested FS regions. In contrast, the dominant CDW, with $\lambda_2\,\simeq 2.8\,a$, cannot be explained within the conventional EPC-assisted Peierls framework. Instead, the observed thermal evolution and the calculated susceptibility suggest that this CDW -- unique to the ML -- arises from a flat region of the electronic band structure, where interactions and correlation effects are expected to dominate. 

\paragraph{Impact}
Monolayer VSe$_{2}$ stands apart in hosting two coexisting charge orders with distinct physical origins. Conventional electronic materials are typecast by the mechanisms and phenomena they host. Our work suggests that ML-VSe$_{2}$ transcends such labelling, and hosts coexisting ordered states originating from contrasting coupling mechanisms. The prospect of such emergent electron correlations and ensuing ordered states presenting themselves in 2D TMDCs more generally is particularly promising given their predominance in the plethora of proposed designs for heterogeneous layered materials~\citep{Novoselov2005b, Geim2013, Novoselov2016}. Their potential for tunability and their interplay with conventional charge and spin orders in the ultrathin limit is promising for realizing exotic ordered states on one the hand, and for applications in multifunctional electronics on the other.

\section{Methods\label{sec:Methods}}

\begin{small}
\paragraph{{\small{}VSe$_{2}$ growth}}
\noindent \textbf{Film Growth.} $1T$-VSe$_{2}$ films were grown on HOPG substrates in a home-made, ultrahigh-vacuum molecular beam epitaxy (MBE) system, the growth chamber of which has a base pressure of $2 \times 10^{-9}$\,mbar. The substrate was exfoliated \emph{ex-situ}, immediately transferred into the MBE chamber, and then outgassed at 420\,$^{\circ}$C for 3\,h before MBE growth. The VSe$_{2}$ samples were grown \emph{via} simultaneously evaporating V and Se using an electron-beam evaporator and a Knudsen cell, respectively, onto the substrates maintained at 360 $^{\circ}$C. The Se/V ratio was high, and Se was controlled to be in excess. A selenium capping layer was deposited onto the VSe$_{2}$ surface to prevent direct ambient contamination during \emph{ex-situ} transport to the varying temperature STM/nc-AFM system for subsequent measurements. The capping layer was removed by annealing at 240\,$^{\circ}$C for 30\,min in the microscope chamber.%

\paragraph{{\small{}STM/nc-AFM measurements}}

\textbf{STM \& AFM Measurements.} STM/nc-AFM measurements were performed over 78-204\,K in an Omicron UHV system interfaced to a Nanonis controller equipped with STM/qPlus sensor and an electrical local heater. To reduce thermal drift during data acquisition, the STM was first allowed to stabilise at each temperature. Electrochemically etched tungsten tips were used with bias voltage applied to the the tip, while the sample holder was grounded. STM images were acquired using constant current mode. For nc-AFM imaging, the constant-height mode with an oscillation amplitude of 10\,nm was used to record the frequency shift (${\Delta}f$) of the qPlus resonator (sensor frequency $f_{0}\approx24$\,kHz, $Q\approx8000$). A lock-in technique was used to measure $dI/dV$ spectra, with a modulation of 625\,Hz and 30\,mV.%

\paragraph{{\small{}DFT Calculations}}
\textbf{DFT Calculations.}  1T-VSe$_{2}$ belongs to the space group $P\bar{3}m1$, with the lattice parameters for the monolayer crystal being $a=b=3.33$\,Å, $\gamma=120^{\circ}$~\citep{Li2014}. First-principle atomic and electronic structure calculations were performed within the density functional theory (DFT) framework as implemented in the Vienna Ab-initio Simulation Package (VASP)~\citep{Kresse1996b} with a plane-wave basis up to a cut-off of 500\,eV.  
To simulate the monolayer, we artificially set the distance between two layers of VSe$_{2}$ to $25$\,Å. The Perdew-Burke-Ernzerhof (PBE)~\citep{Perdew1996a} form was used for the exchange-correlation functional. The $\Gamma$-centred $k$-mesh was set to be $25\times 25\times1$ in the Brillouin zone for the self-consistent calculation. The relaxation of atomic structure was done in two ways. First, a $4\times4$ supercell was relaxed under the symmetry constraints of the space group. This process was then repeated without any symmetry constraints applied.

\paragraph{{\small{}Tight Binding Calculations}}
\textbf{Tight-Binding Calculations.}  A tight-binding fit was performed for the single $d$-orbital band crossing the Fermi level in the available ARPES data of~\citep{Feng2018}. To obtain the best fit, we used an expansion of the dispersion $E(\mathbf{k})$ in functions respecting the lattice symmetries. Including terms to fifth order, the fit can be expressed as:
\begin{align}
\begin{split}\label{eq:H_TB}E(\mathbf{k})= & \; t_{0}+t_{1}\Big(2\cos(\xi)\cos(\eta)+\cos(2\xi)\Big)\\
 & +t_{2}\big(2\cos(3\xi)\cos(\eta)+\cos(2\eta)\big)\\
 & +t_{3}\big(2\cos(2\xi)\cos(2\eta)+\cos(4\xi)\big)\\
 & +t_{4}\big(\cos(\xi)\cos(3\eta)+\cos(5\xi)\cos(\eta)\\
 & \qquad +\cos(4\xi)\cos(2\eta)\big)\\
 & +t_{5}\big(2\cos(3\xi)\cos(3\eta)+\cos(6\xi)\big),
\end{split}
\end{align}
where $\xi=\frac{1}{2}k_{x}$ and $\eta=\frac{\sqrt{3}}{2}k_{y}$, while $k_{x},k_{y}$ are given in units of $\frac{2\pi}{a}$, with $a$ the lattice parameter. $t_{i}$ are the (in-plane) hopping amplitudes. The best fit to ARPES data based on this form of the dispersion is shown in \ref{fig:BandStructure}.

\paragraph{{\small{Susceptibility Calculations}}}
\textbf{Susceptibility Calculations.} In the limit of weak electron-phonon coupling, the electronic susceptibility can be derived from a perturbative expansion of the phonon propagator, using the random phase approximation (RPA). Neglecting vertex corrections, which should be small~\citep{Migdal1958}, the renormalised phonon propagator is described by $D_{RPA}=(D_{0}^{-1}-D_{2})^{-1}$, with bare phonon propagator $D_{0}$ and electronic susceptibility $D_{2}$, given by~\citep{Doran1978a,Flicker2015,Henke2020}: 
\begin{align}
D_{2}(\mathbf{q})=-\sum_{\mathbf{k}\in\text{BZ}}|g_{\mathbf{k},\mathbf{k+q}}|^{2}\frac{f(E_{\mathbf{k}})-f(E_{\mathbf{k+q}})}{E_{\mathbf{k}}-E_{\mathbf{k+q}}+i\delta}.
\end{align}
Here, $f(E)$ is the Fermi-Dirac distribution function, $E_{\mathbf{k}}$ is the non-renormalised electronic dispersion and we use a small regulator $\delta=0.1$\,meV. If the system has an intrinsic or electron-phonon driven CDW instability within the weak-coupling limit, the susceptibility will exhibit a maximum at the CDW wave vector $\mathbf{Q}$.
Generally, the electron-phonon coupling (EPC) matrix elements $|g_{\mathbf{k},\mathbf{k+q}}|^{2}$ are difficult to compute exactly. For this reason, it is common to set them to unity, resulting in the Lindhard function: 
\begin{align}
\chi(\mathbf{q})=-\sum_{\mathbf{k}\in\text{BZ}}\frac{f(E_{\mathbf{k}})-f(E_{\mathbf{k+q}})}{E_{\mathbf{k}}-E_{\mathbf{k+q}}+i\delta}.
\end{align}
In previous works, it has been shown that the EPC matrix elements can be approximated based purely on the electronic dispersion. This approximation has been well-tested for transition metal compounds with $d$-orbital character at $E_{F}$~\citep{Varma1979,Flicker2016,Henke2020}. In the case of a single band crossing $E_{F}$, the expression becomes: 
\begin{equation}
\mathbf{g}_{\mathbf{k},\mathbf{k+q}}\propto\frac{\partial E_{\mathbf{k}}}{\partial\mathbf{k}}-\frac{\partial E_{\mathbf{k+q}}}{\partial\mathbf{k}}.
\end{equation}
The orientation of $\mathbf{g}_{\mathbf{k},\mathbf{k+q}}$ indicates the direction of phonon polarisation. We consider longitudinal CDWs, such that the relevant component of the EPC vector is parallel to the in-plane phonon momentum: ${g}_{\mathbf{k},\mathbf{k+q}}=\mathbf{g}_{\mathbf{k},\mathbf{k+q}}\cdot\mathbf{q}_{\parallel}/|\mathbf{q}_{\parallel}|$. %

\end{small}

\noindent \begin{center}
{\small{}\rule[0.5ex]{0.4\columnwidth}{0.5pt}}{\small\par}
\par\end{center}
\vspace{1ex}
\added{\noindent \textsf{\textbf{Supporting Information.}} Supporting data on MoS$_2$ substrate, comparison of STM- and AFM-imaging of ML-VSe$_2$, full dataset for temperature-dependence presented in manuscript, and comparison of tight-binding fit to ARPES data.}

\vspace{1ex}
\added{\noindent \textsf{\textbf{Author Contributions.}} R.C., Y.L.H., J.G., and X.H. performed the experiments and analysed the data. S.S. and T.D. performed the ab-initio calculations. J.H. performed the tight-binding and susceptibility calculations. J.v.W., A.S., and A.T.S.W. coordinated and supervised the work. All authors discussed the results and provided inputs to the manuscript.}

\vspace{1ex}

\noindent \textsf{\textbf{Financial Interests.}} The authors declare no competing financial interests.
\vspace{1ex}

\noindent \textsf{\textbf{Acknowledgments.}} The work in Singapore was funded by the Ministry of Education Tier 2 grant no. MOE2017-T2-2-139. J.G. acknowledges funding from NRF-NSFC under grant no. R-144-000-405-281, and A.S. acknowledges the support of A*STAR Singapore under the SpOT-LITE programme (Grant no. A18A6b0057).
T.D. acknowledges the support of the Department of Science, Government of India under the project R(RO)/DST/NSM/HPC\_Applications/2021-1464. The computational work at IISc also benefited from the S.E.R.C. supercomputing facility. 
%

\bibliography{VSe2-CDW_Refs}

\noindent \begin{center}
{\small{}\rule[0.5ex]{0.6\columnwidth}{0.5pt}}{\small\par}
\par\end{center}


\setcounter{secnumdepth}{4}
\setcounter{figure}{0}
\renewcommand\thesection{S\arabic{section}}
\renewcommand{\theparagraph}{S\arabic{section}\alph{paragraph}}
\makeatletter\@addtoreset{paragraph}{section}\makeatother
\makeatletter\def\p@paragraph{}\makeatother
\renewcommand{\thefigure}{S\arabic{figure}}
\renewcommand{\theequation}{S\arabic{equation}}
\renewcommand{\thetable}{S\arabic{table}}
\newcommand{\cmark}{\textcolor{red}{\ding{51}}}%
\newcommand{\xmark}{\ding{55}}%

\clearpage{}
\onecolumngrid

\begin{center}\Large{\textbf{Supplemental Information} }
\end{center}
\section{STM data on M\lowercase{o}S$_2$}
\begin{figure}[h]
    \begin{centering}
    \includegraphics[width=0.8\linewidth]{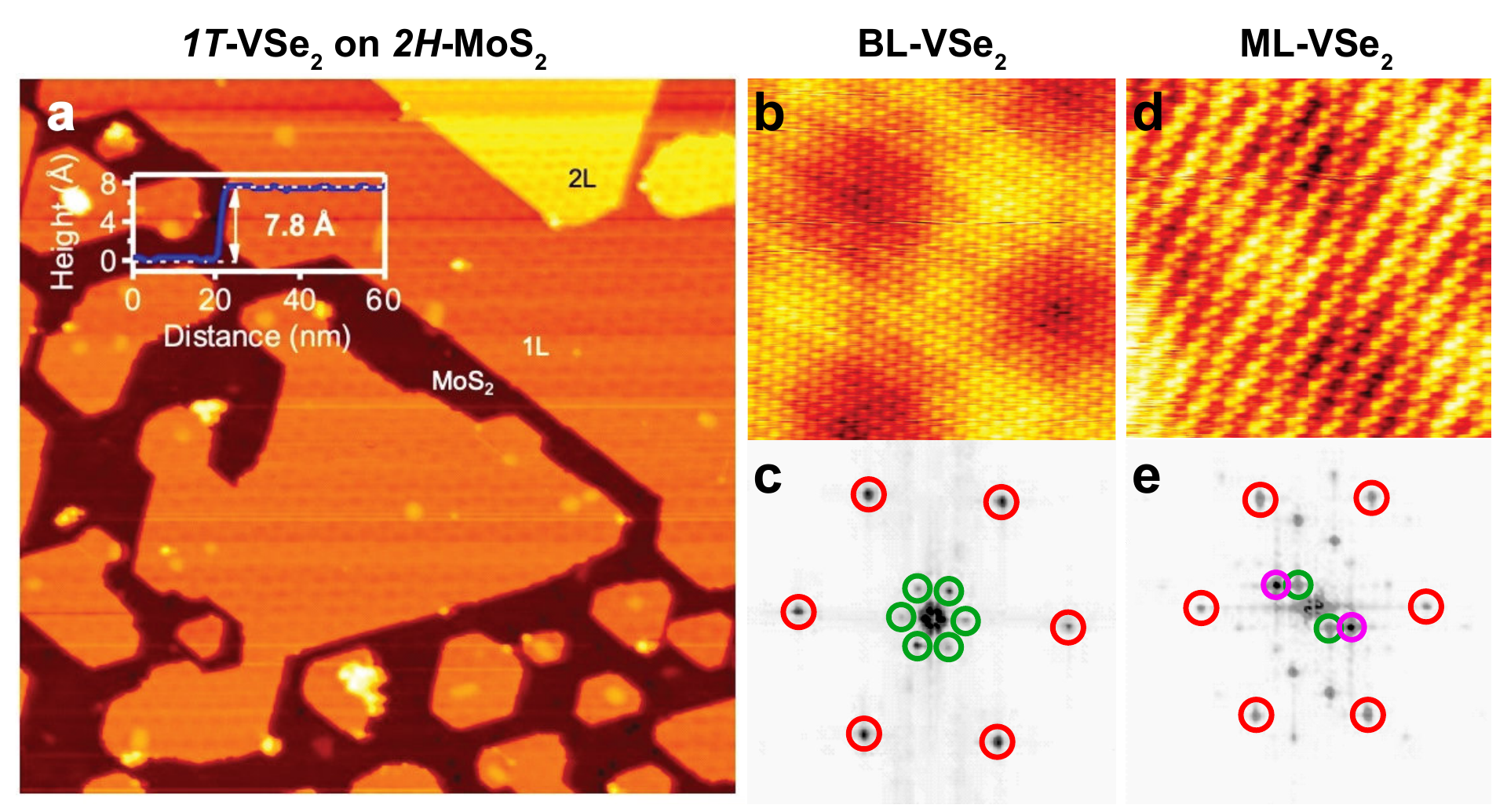}
    \end{centering}
    \caption[VSe$_2$ on MoS$_2$]{\textbf{STM Imaging of Ultrathin VSe$_2$ on MoS$_2$ Substrate.} 
    (a) Large-scale STM topograph ($200 \times200$\,nm$^2$; $V_{\rm tip} = 2.4$\,V, $I_{\rm tip} = 100$\,pA) of ultrathin $1T$-VSe$_2$ grown on $2H$-MoS$_2$. Inset: line profile showing that the first VSe$_2$ layer is ~$7.8$\,\AA~above the substrate. (b-e) Atomic resolution low-temperature) STM topographs ($10\times10$\,nm$^2$) of BL- (b) and ML-VSe$_2$ (c) at $78$\,K, and their respective Fourier transforms (FTs). $I_{\rm tip} = 180$\,pA, (b, d)$V_{\rm tip} = 10$\,mV (b), -10\,mV (d). \label{fig:SI_mos2}}
\end{figure}
\vspace{10pt}

The STM data reported in the manuscript is acquired on VSe$_2$ grown on a HOPG substrate. Samples were grown similarly (by MBE, see Methods) on a $2H$-MoS$_2$ substrate~\citep{Chua2020}. A large-scale STM topograph, shown in \ref{fig:SI_mos2}a, indicates that the first VSe$_2$ layer lies 7.8\,\AA~above the substrate. 
The remaining panels in \ref{fig:SI_mos2} show atomic resolution zoom-ins of defect-free regions corresponding to BL- and ML-VSe$_2$. The slight lattice mismatch between VSe$_2$ and the substrate generates a hexagonal moir\'{e} superstructure, which is clearly visible in the topographs (Fig. \ref{fig:SI_mos2}b-c). The new supercell consists of $17 \times  17$ VSe$_2$ unit cells atop $18 \times 18$ MoS$_2$ unit cells~\citep{Chua2020}. In contrast, no moir\'{e} patterns was observed for ultrathin VSe$_2$ on HOPG for any setpoint. Meanwhile, the Fourier transforms (FTs) of the zoomed-in topographs demonstrate, however, that aside from the difference in moiré intensity, the same CDW peaks are present for VSe$_2$ grown on  $2H$-MoS$_2$ as seen for VSe$_2$ grown on HOPG (see manuscript Fig. 1-2). That is, the BL hosts a triple-$Q$, $4a$ CDW phase just like the bulk (green circles in \ref{fig:SI_mos2}c), while the ML hosts two unidirectional CDWs with $Q_1=0.25$\,a* and $Q_2=0.36$\,a* (green and magenta circles, respectively). 

\clearpage

\begin{center}\Large{\textbf{Supplemental Information} }
\end{center}
\section{Comparison of STM and AFM Imaging of ML-VS\lowercase{e}$_2$}
\begin{figure}[h]
    \begin{centering}
    \includegraphics[width=0.5\linewidth]{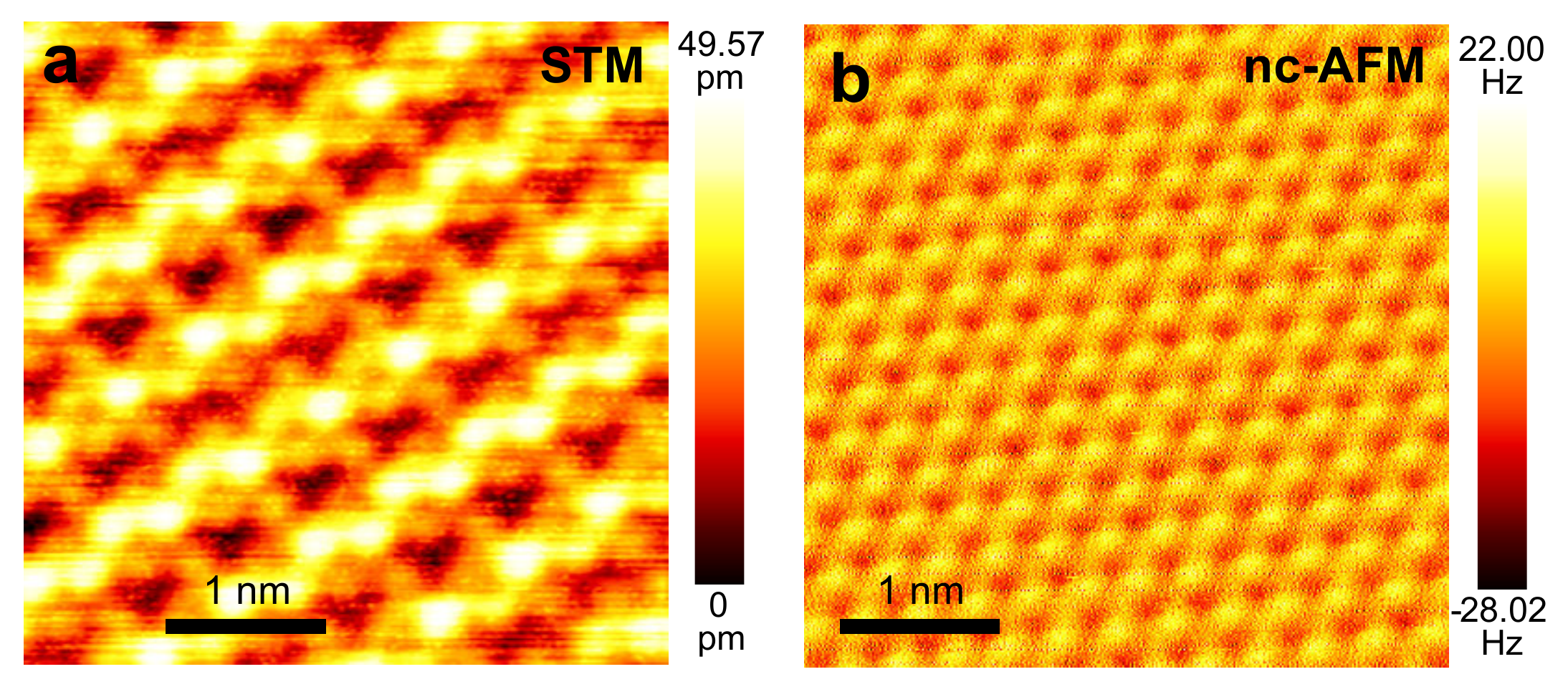}
    \end{centering}
    \caption[nc-AFM on ML-VSe$_2$]{\textbf{STM and AFM topographs of ML-VSe$_2$.} 
    Comparison of atomic resolution topographs ($4\times4$\,nm$^2$) of ML-VSe$_2$ on HOPG at $T = 78$~K acquired by \textbf{(a)} STM ($V_{\rm tip}=-0.7$\,V, $I_{\rm tip} = 220$\,pA), and \textbf{(b)} non-contact AFM techniques. Superstructures are visible in (a), but not in (b).
    \label{fig:SI_afm}} 
\end{figure}
\vspace{10pt}

In ref.~\citep{Duvjir2018}, it was suggested that the additional superstructures seen in STM FTs beyond $Q_1$ (identified in our case as $Q_2$) was purely due to structural distortions of the atomic lattice. If this is the case, one would expect the deformation to also be visible in non-contact atomic force microscopy (nc-AFM) topographs acquired under similar sample conditions as STM (see Methods). As we demonstrate in \ref{fig:SI_afm}, however, ac-AFM measurements at 78 K show ML-VSe$_2$ to be atomically flat within the experimental resolution. This, in combination with the thermal rotation of the orientation of $Q_2$ , and the lack of evidence for structural instabilities in our \textit{ab-initio} atomic relaxation (see main text), allow us to interpret $Q_2$ as a charge density wave (CDW). Meanwhile, the expected magnitude of the Peierls distortion associated with CDWs is $\sim 10^{-2}\,a$, \emph{i.e.} well below the resolution of available microscopy techniques.  

\clearpage

\begin{center}\Large{\textbf{Supplemental Information} }
\end{center}
\section{Temperature-Dependent STM Data of BL- and ML-VS\lowercase{e}$_2$}
\begin{figure}[h]
    \begin{centering}
    \includegraphics[width=\linewidth]{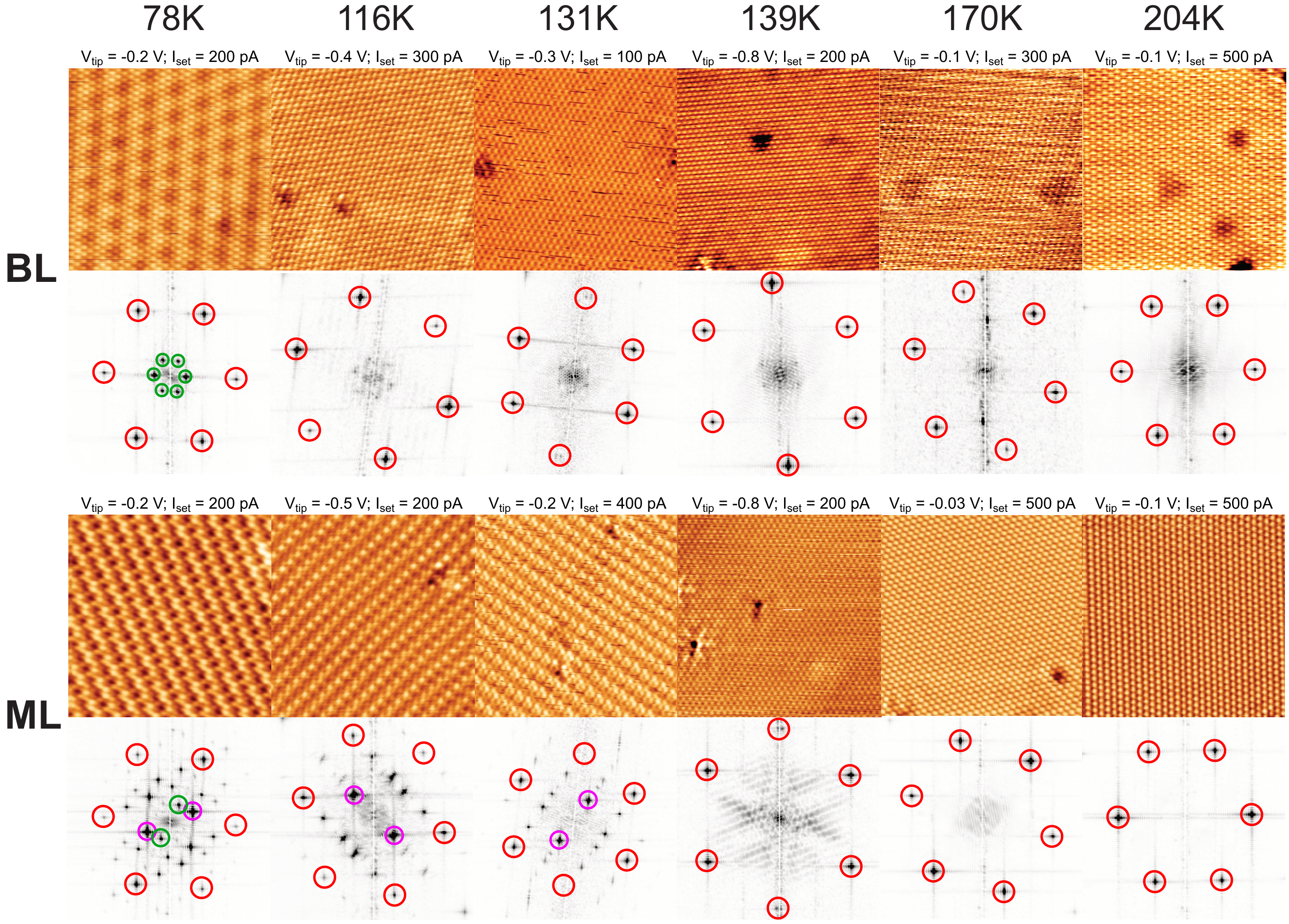}
    \end{centering}
    \caption[VSe$_2$ on HOPG, T-dependence]{\textbf{Temperature-dependent STM imaging of BL- and ML-VSe$_2$.} Atomically resolved STM topographs (approx. $10 \times 10$~nm$^2$) and their respective FTs of BL- (top half) and ML-VSe$-2$ (bottom half), acquired at temperatures varying over 78-204 K. Imaging conditions for each topograph are indicated. The data shown is used in Fig. 3 of the main text. 
    \label{fig:SI_STM-T-dep}}
\end{figure}
\vspace{10pt}

\clearpage

\begin{center}\Large{\textbf{Supplemental Information} }
\end{center}
\section{Comparing Tight-Binding Fit to ARPES}
\begin{figure}[h]
    \begin{centering}
    \includegraphics[width=0.4\linewidth]{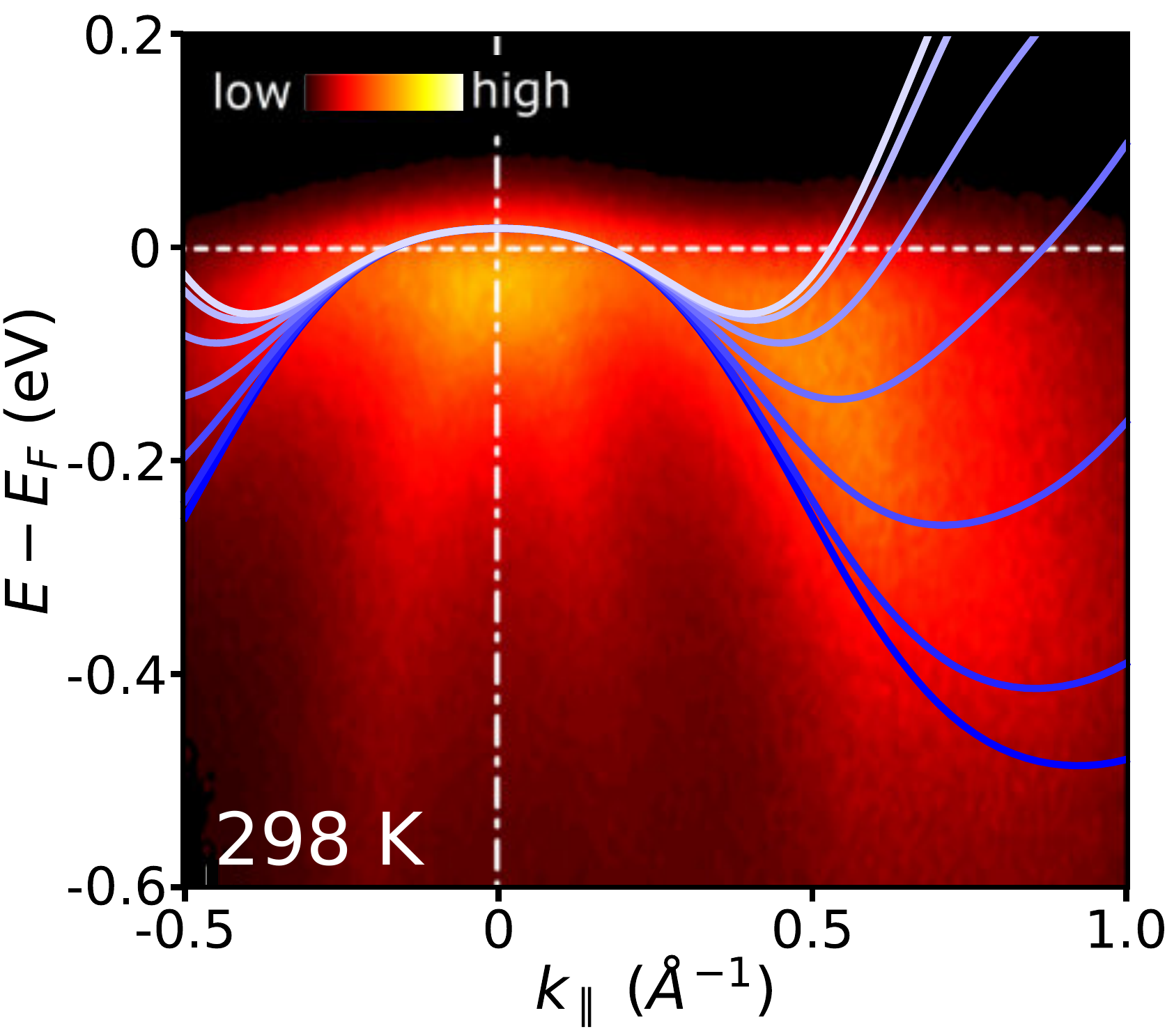}
    \end{centering}
    \caption[TB vs. ARPES of same samples.]{\textbf{Comparing TB fit to ARPES spectrum.} Azimuthally averaged ARPES spectrum acquired on an ultrathin VSe$_2$ sample grown on HOPG, reported previously~\citep{Wong2019}. The sample is from the same batch as that used for the STM measurements reported in the manuscript. Lines show TB fits for a sequence of $\mathbf{k}$-directions in steps of $10^\circ$, from along $\Gamma-\mathrm{M}$ (dark blue) to $\Gamma-\mathrm{K}$ (white).\label{fig:SI_TB-ARPES}}
\end{figure}
\vspace{10pt}

The tight-binding (TB) model used in the susceptibility calculations (main text) was fitted using a least-squares fit procedure to the peak positions of the ARPES spectrum of ML-VSe$_2$ on bilayer graphene reported in ref.~\citep{Feng2018}. To ensure that the fitted spectrum is also consistent with the samples used in the present study, we compare our tight-binding fit to the ARPES spectrum of a sample from the same batch of ML-VSe$_2$ grown on HOPG, previously reported in ref.~\citep{Wong2019}. As the macroscopic ARPES beam averages over ML-VSe$_2$ grains of multiple orientations, the resulting spectrum corresponds to an azimuthal average over $k(x,y)$-space. While this limits a direct experimental determination of the $\bf{k}$-resolved Fermi surface for the samples studied by STM, here we compare the tight-binding fit band structure with the azimuthally averaged ARPES data. To account for the azimuthal averaging of this spectrum, we plot the tight-binding spectrum (\ref{fig:SI_TB-ARPES}) for a sequence of different $\mathbf{k}$-directions, rotated $10^\circ$ from one another, lying along $\Gamma-\mathrm{M}$ (dark blue) to $\Gamma-\mathrm{K}$ (white). As can be seen in the figure, the tight-binding spectra and ARPES spectra are in good agreement.

\end{document}